\documentclass[12pt]{article}
\usepackage{amsmath} 
\usepackage{amssymb}
\usepackage{hhline}
\usepackage[scale={.75,.75}]{geometry}
\usepackage{url}
\usepackage{lscape}
\usepackage{caption}
\usepackage[numbers,sort&compress]{natbib}
\usepackage{epsfig}
\usepackage{multirow} 
\usepackage{color}
\usepackage{verbatim}
\usepackage{nicefrac}
\usepackage{upgreek}
\usepackage{bbm}
\usepackage{setspace}
\usepackage{pstricks}
\usepackage{psfrag}
\usepackage{color}

\newcommand{\slashed}[1]{\displaystyle{\not}{#1}}

\usepackage{hyperref}

\newcommand{\p}{\textbf{p}}
\definecolor{green}{rgb}{0,0.5,0}  

\begin{document}
\date{}

\title{\vspace{-1.5cm} 
\begin{flushright}
\vspace{-0.4cm}
{\scriptsize \tt TUM-HEP-1049/16}  
\end{flushright}
{\bf Neutrinoless double $\beta$ decay and low scale leptogenesis
}
}

\author{
\vspace{0.2cm}
Marco Drewes$^a$, Shintaro Eijima$^b$ 
\\
\footnotesize{$^a$Physik Department T70, Technische Universit\"at M\"unchen,}\\
\footnotesize{ James Franck Stra\ss e 1, D-85748 Garching, Germany}\\
\footnotesize{$^b$Institute of Physics,
\'Ecole Polytechnique F\'ed\'erale de Lausanne,}\\
\footnotesize{CH-1015 Lausanne, Switzerland}
}
\maketitle

\begin{abstract}
  \noindent The extension of the Standard Model by right handed neutrinos with masses in the GeV range can simultaneously explain the observed neutrino masses via the seesaw mechanism and the baryon asymmetry of the universe via leptogenesis. It has previously been claimed that the requirement for successful baryogenesis implies that the rate of neutrinoless double $\beta$ decay in this scenario is always smaller than the standard prediction from light neutrino exchange alone. In contrast, we find that the rate for this process can also be enhanced due to a dominant contribution from heavy neutrino exchange. In a small part of the parameter space it even exceeds the current experimental limit, while the properties of the heavy neutrinos are consistent with all other experimental constraints and the observed baryon asymmetry is reproduced.  This implies that neutrinoless double $\beta$ decay experiments have already started to rule out part of the leptogenesis parameter space that is not constrained by any other experiment, and the lepton number violation that is responsible for the origin of baryonic matter in the universe may be observed in the near future.
  \end{abstract}



\section{Introduction}\label{sec:introduction}
With the exception of neutrinos, all fermions in the Standard Model (SM) of particle physics are known to exist with both left handed (LH) and right handed (RH) chirality. If RH neutrinos exist, they can explain the observed neutrino flavour oscillations via the seesaw mechanism \cite{Minkowski:1977sc,GellMann:seesaw,Mohapatra:1979ia,Yanagida:1980xy,Schechter:1980gr,Schechter:1981cv}. In addition, RH neutrinos may also explain the baryon asymmetry of the universe (BAU) \cite{Canetti:2012zc} via leptogenesis during their CP violating decays \cite{Fukugita:1986hr} or CP violating oscillations \cite{Akhmedov:1998qx,Asaka:2005pn} in the early universe, or compose the Dark Matter (DM) \cite{Adhikari:2016bei}. 
In Refs.~\cite{Asaka:2005pn,Asaka:2005an} it has been proposed that all of these puzzles can be solved simultaneously by RH neutrinos alone, which was found to be feasible in Refs.~\cite{Canetti:2012vf,Canetti:2012kh}. A pedagogical review of this scenario, which is known as the \emph{Neutrino Minimal Standard Model} ($\nu$MSM), can be found in Ref.~\cite{Boyarsky:2009ix}.
Finally, light RH neutrinos could also act as Dark Radiation in the early universe and explain the observed neutrino oscillation anomalies \cite{Abazajian:2012ys}. A general review on the role of RH neutrinos in particle physics and cosmology can e.g.\ be found in Ref.~\cite{Drewes:2013gca}.
In the present work we focus on the possibility that RH neutrinos $N_I$ with Majorana masses $M_I$ in the GeV range can simultaneously explain the observed neutrino oscillations and the baryon asymmetry of the universe without violating any of the known experimental or cosmological constraints on their properties \cite{Atre:2009rg,Antusch:2014woa,Drewes:2015iva,deGouvea:2015euy,Fernandez-Martinez:2016lgt}.

Experimentally the GeV range is very interesting because the RH neutrinos can be searched for in meson decays at b-factories \cite{Canetti:2014dka,Milanes:2016rzr} or fixed target experiments \cite{Gorbunov:2007ak}, including NA62 \cite{Asaka:2012bb}, the SHiP experiment proposed at CERN \cite{Anelli:2015pba,Alekhin:2015byh,Graverini:2015dka} or a similar setup proposed at the DUNE beam at FNAL \cite{Akiri:2011dv,Adams:2013qkq}.
With sufficient statistics, it might even be possible to measure the CP violation in the $N_I$ decay \cite{Cvetic:2015naa}.
Theoretically the \emph{low scale seesaw} is motivated by models based on classical scale invariance \cite{Khoze:2013oga}, 
in the framework of the ``inverse seesaw'' 
\cite{Mohapatra:1986bd,Mohapatra:1986aw} and other models with an approximate conservation of lepton number (e.g. \cite{Chikashige:1980ui,Gelmini:1980re,Wyler:1982dd,GonzalezGarcia:1988rw,Branco:1988ex,Abada:2007ux,Shaposhnikov:2006nn,Gavela:2009cd,Sierra:2012yy,Racker:2012vw,Fong:2013gaa}) or by applying Ockham's razor to the number of new particles required to explain the known beyond the SM phenomena~\cite{Asaka:2005pn}. 
Placing the seesaw scale in the GeV range can avoid the hierarchy problem of the Higgs mass \cite{Shaposhnikov:2007nj}, to which superheavy RH neutrinos would contribute \cite{Vissani:1997ys}, while avoiding cosmological constraints that disfavour heavy neutrino masses below 100 MeV \cite{Hernandez:2014fha}.

It has been pointed out by different authors \cite{Asaka:2011pb,LopezPavon:2012zg,Lopez-Pavon:2015cga,Gorbunov:2014ypa,Drewes:2015iva} that the rate for neutrinoless double $\beta$ decay in the presence of RH neutrinos with GeV masses can significantly differ from the standard prediction from light neutrinos alone. 
In this work we address the question whether an large rate of neutrinoless double $\beta$ decay can be realised while simultaneously generating the observed BAU. Previous studies have found that this requirement suppresses the rate of neutrinoless double $\beta$ decay \cite{Bezrukov:2005mx,Asaka:2011pb,Asaka:2013jfa}.
A key point in the line of argument was the assumption that a degeneracy in the heavy neutrino masses is required for leptogenesis if they lie in the GeV range. 
However, the mass degeneracy is not a necessary requirement for low scale leptogenesis if there are more than two heavy neutrinos \cite{Drewes:2012ma}.

In this letter we show the rate of neutrinoless double beta decay in the scenario with three RH neutrinos can exceed that only from light neutrino exchange while explaining the BAU via leptogenesis. 
Furthermore we show in a numerical parameter scan that even in the scenario with two RH neutrinos, which is the minimal number to explain the observed neutrino oscillations, there exists a corner in parameter space in which this is possible. 

\section{The seesaw model}
The (type I) seesaw model is defined by adding $n$ RH neutrinos $\nu_R$ to the SM, which leads to the Lagrangian
\begin{eqnarray}
	\label{L}
	\mathcal{L} &=&\mathcal{L}_{SM}+ 
	i \overline{\nu_R}\slashed{\partial}\nu_R-
	\overline{\ell_{L}}F\nu_{R}\tilde{\Phi} -
	\tilde{\Phi}^{\dagger}\overline{\nu_{R}}F^{\dagger}\ell_L 
-{\rm \frac{1}{2}}(\overline{\nu_R^c}M_{M}\nu_{R} 
	+\overline{\nu_{R}}M_{M}^{\dagger}\nu^c_{R}). 
	\end{eqnarray}
$\mathcal{L}_{SM}$ is the SM Lagrangian, 
$\ell_{L}=(\nu_{L},e_{L})^{T}$ are the SM lepton doublets and $\Phi$ is the Higgs doublet with $\tilde{\Phi}=\epsilon\Phi^*$.
Here $\epsilon$ is the antisymmetric $SU(2)$-invariant tensor. 
$M_{M}$ a Majorana mass term for $\nu_{R}$  and $F$ is a matrix of Yukawa couplings.
We have defined $\nu_R^c\equiv C\overline{\nu_R}^T$, where the charge conjugation matrix is $C=i\gamma_2\gamma_0$. We work in the heavy neutrino mass basis in flavour space, i.e., $(M_M)_{IJ}=\delta_{IJ}M_I$.
Adding $n$ RH neutrinos to the SM introduces $7n-3$ new physical parameters.
The relation between these parameters
and the parameters constrained by neutrino oscillation data \cite{Gonzalez-Garcia:2015qrr} can be 
expressed in terms of the Casas-Ibarra parametrisation \cite{Casas:2001sr}
\begin{align}\label{CasasIbarraDef}
F=\frac{i}{v}U_\nu\sqrt{m_\nu^{\rm diag}}\mathcal{R}\sqrt{M^{\rm diag}}\,
\end{align}
with $(m_\nu^{\rm diag})_{ij}=\delta_{ij} m_i$, where $m_i$ are the light neutrino masses.
The matrix $U_\nu$ can be factorised as
\begin{align}
\label{PMNS}
U_\nu=V^{(23)}U_\delta V^{(13)}U_{-\delta}V^{(12)}{\rm diag}(e^{i \alpha_1/2},e^{i \alpha_2 /2},1)\,,
\end{align}
with $U_{\pm \delta}={\rm diag}(e^{\mp i \delta/2},1,e^{\pm i \delta /2})$.
The non vanishing entries of the matrix $V=V^{(23)}V^{(13)}V^{(12)}$ are given by:
\begin{align}
V^{(ij)}_{ii}=V^{(ij)}_{jj}=\cos \uptheta_{ij} \ , \
V^{(ij)}_{ij}=-V^{(ij)}_{ji}=\sin \uptheta_{ij} \ ,\
V^{(ij)}_{kk}=1 \quad \text{for $k\neq i,j$}.
\end{align}
The parameters $\uptheta_{ij}$ are the light neutrino mixing angles, $\delta$ is referred to as the Dirac phase and $\alpha_{1,2}$ as Majorana phases.
The complex orthogonal matrix $\mathcal{R}$  fulfils the condition $\mathcal{R}\mathcal{R}^T=1$. In case of $n=3$ it can be expressed as
\begin{align}
\mathcal{R}=\mathcal{R}^{(23)}\mathcal{R}^{(13)}\mathcal{R}^{(12)}
\end{align}
where the non-vanishing entries are given by the three complex ``Euler angles'' $\omega_{ij}$,
\begin{align}
\mathcal{R}^{(ij)}_{ii}=\mathcal{R}^{(ij)}_{jj}=\cos \omega_{ij} \ , \
\mathcal{R}^{(ij)}_{ij}=-\mathcal{R}^{(ij)}_{ji}=\sin \omega_{ij} \ , \
\mathcal{R}^{(ij)}_{kk}=1 \quad \text{for $k\neq i,j$}.
\end{align}
For two flavours there is only one complex angle $\omega$, and one has to distinguish between normal ordering (NO) and inverted ordering (IO):
\begin{align}
\mathcal{R}^{\rm NO}=
\begin{pmatrix}
0 && 0\\
\cos \omega && \sin \omega \\
-\xi \sin \omega && \xi \cos \omega
\end{pmatrix}\,,\quad \quad 
\mathcal{R}^{\rm IO}=
\begin{pmatrix}
\cos \omega && \sin \omega \\
-\xi \sin \omega && \xi \cos \omega \\
0 && 0
\end{pmatrix}
\,,
\end{align}
where $\xi=\pm 1$. 
When the Higgs field obtains an expectation value $v(T)$, the Yukawa couplings lead to mixing between $\nu_R$ and $\nu_L$. 
This mixing can be quantified by the matrix
\begin{equation}
\theta=v  F M_M^{-1}.
\end{equation}
In general, the mass eigenstates can be expressed in terms of the Majorana spinors
\begin{equation}\label{LightMassEigenstates}
\upnu_i=\left[ V_\nu^{\dagger}\nu_L-U_\nu^{\dagger}\theta\nu_{R}^c + V_\nu^{T}\nu_L^c-U_\nu^{T}\theta\nu_{R} \right]_i
\end{equation}
 which can be identified with the light neutrinos with masses $m_i$, 
and 
\begin{equation}
N_I=\left[V_N^\dagger\nu_R+\Theta^{T}\nu_{L}^{c} +  V_N^T\nu_R^c+\Theta^{\dagger}\nu_{L}\right]_I.
\end{equation}
The observed light mass eigenstates $\upnu_i$ are connected to the active flavour eigenstates by the matrix $V_\nu$, which is related to $U_\nu$ via 
$V_\nu= (\mathbbm{1}-\frac{1}{2}\theta\theta^{\dagger})U_\nu$. 
$V_N$ and $U_N$ are their equivalents in the sterile sector; $U_N$ diagonalises the heavy neutrino mass matrix $M_N=M_M + \frac{1}{2}(\theta^{\dagger} \theta M_M + M_M^T \theta^T \theta^{*})$ after electroweak symmetry breaking, and $V_N= (1-\frac{1}{2}\theta^T\theta^*)U_N$. 
The mixing between the heavy and light states can is finally given by 
\begin{equation}
\Theta_{\alpha I}=(\theta U_N^*)_{\alpha I}.
\end{equation}
The overall magnitude of the mixing is governed by the
imaginary part of the complex angels $\omega$ or $\omega_{ij}$.
For instance, for $n=2$ one finds 
\begin{eqnarray}
{\rm tr}[\Theta^\dagger \Theta]&=&\frac{M_2-M_1}{2M_1 M_2} (m_2-m_3)\cos(2 {\rm Re}\omega)+\frac{M_1+M_2}{2M_1 M_2}(m_2+m_3)\cosh(2 {\rm Im}\omega) 
\end{eqnarray}
with normal ordering and
\begin{eqnarray}
{\rm tr}[\Theta^\dagger \Theta]&=&\frac{M_2-M_1}{2M_1 M_2} (m_1-m_2)\cos(2 {\rm Re}\omega)+\frac{M_1+M_2}{2M_1 M_2}(m_1+m_2)\cosh(2 {\rm Im}\omega) 
\end{eqnarray}
with inverted ordering.

\section{Neutrinoless double $\beta$ decay}
\paragraph{General case} - In the context of neutrino physics, constraints on the lifetime of neutrinoless double $\beta$ decay are commonly expressed in terms of the quantity
\begin{equation}\label{mee}
m_{\beta\beta}=\left|
\sum_i (U_\nu)_{ei}^2m_i + \sum_I \Theta_{eI}^2M_I f_A(M_I)
\right|.
\end{equation}
The first term is the contribution due to the exchange of light neutrinos,
\begin{equation}
m_{\beta\beta}^\nu=\sum_i (U_\nu)_{ei}^2m_i.
\end{equation}
The second term comes from heavy neutrino exchange. For $M_I$ larger than the typical momentum exchange $\sim100$ MeV in neutrinoless double $\beta$ decay, the $N_I$ are virtual. The suppression due to this virtuality is parametrised by the function $f_A$, which suffers from some uncertainty due to uncertainties in the nuclear matrix elements that determine the exchanged momentum. For our purpose, we approximate it by
\begin{equation}
f_A(M)\simeq \frac{\Lambda^2}{\Lambda^2+M^2}\Big|_{\Lambda^2= (0.159{\rm GeV})^2},
\end{equation}
which corresponds to the ``Argonne'' model discussed in Ref.~\cite{Faessler:2014kka}. 
Here $\Lambda$ is the typical momentum exchange in the decay.
At tree level,\footnote{Loop corrections are e.g.\ discussed in Refs.~\cite{Fernandez-Martinez:2015hxa,Drewes:2015iva}.} 
we can use the unitarity relation 
\begin{eqnarray}\label{SeesawConsistency}
\sum_i m_i (U_\nu)_{\alpha i}^2 + \sum_I M_I \Theta_{\alpha I}^2 = 0 
\end{eqnarray}
to rewrite (\ref{mee}) as
\begin{eqnarray}
m_{\beta\beta}&=&\left|
m_{\beta\beta}^\nu
+f_A(\bar{M})\sum_IM_I\Theta_{e I}^2
+\sum_IM_I\Theta_{e I}^2[f_A(M_I)-f_A(\bar{M})]
\right|\nonumber\\
&=&
\left|
[1-f_A(\bar{M})]m_{\beta\beta}^\nu
+\sum_IM_I\Theta_{e I}^2[f_A(M_I)-f_A(\bar{M})]
\right|,\label{meerewritten}
\end{eqnarray}
where $\bar{M}$ is an arbitrarily chosen mass scale.
It is usually assumed that the contribution from $N_I$-exchange is negligible due to the suppression by the function $f_A$. 
Recently several authors have pointed out that this suppression is not efficient enough for $M_I$ in the GeV range \cite{Bezrukov:2005mx,LopezPavon:2012zg,Asaka:2011pb,Asaka:2013jfa,Lopez-Pavon:2015cga,Gorbunov:2014ypa,Drewes:2015iva}, and that the exchange of $N_I$ may dominate neutrinoless double $\beta$ decay. This can significantly modify the allowed regions in the $m_{\rm lightest}$-$m_{\beta\beta}$ plane, which are based on the approximation $m_{\beta\beta}=m_{\beta\beta}^\nu$. 
Here $m_{\rm lightest}$ is the mass of the lightest neutrinos.
So far it has been argued that this can only suppress the rate of neutrinoless double $\beta$ decay in models where the $N_I$ generate the BAU via leptogenesis because it was assumed that successful leptogenesis requires a degeneracy in the heavy neutrino masses \cite{Bezrukov:2005mx,Asaka:2013jfa,Asaka:2011pb,Gorbunov:2014ypa}. Indeed, if the difference $f_A(M_I)-f_A(\bar{M})$ is negligible,  Eq.~(\ref{meerewritten}) reduces to 
\begin{eqnarray}
m_{\beta\beta}\simeq \left|
[1-f_A(\bar{M})]m_{\beta\beta}^\nu
\right|,
\end{eqnarray}
which is always smaller than $m_{\beta\beta}^\nu$.\footnote{The possibility to reduce $m_{\beta\beta}$ below $m_{\beta\beta}^\nu$ is interesting because it means that even a non-observation of neutrinoless double $\beta$ decay at the level $m_{\beta\beta}<10^{-2}$ eV may not rule out the inverted ordering.}
However, it has recently been pointed out \cite{Drewes:2012ma} and confirmed \cite{Shuve:2014zua,Hernandez:2015wna} that the need for a mass degeneracy is specific to the scenarios with $n=2$ and that for $n>2$, leptogenesis from neutrino oscillations does not require a mass degeneracy.

\paragraph{The case $n=2$} - Moreover, one may wonder whether the mass degeneracy of order $10^{-3}$ that is required in the model with $n=2$ is sufficient to suppress the term $\sum_IM_I\Theta_{e I}^2[f_A(M_I)-f_A(\bar{M})]$ in Eq.~(\ref{meerewritten}) for $M_I$ moderately larger than 100 MeV. In absence of a strong mass degeneracy, this term can either increase or reduce $m_{\beta\beta}$.
In the case $n=2$, $m_{\beta\beta}$ can be expressed in terms of the model parameters as
\begin{eqnarray}
m_{\beta\beta}&=&
 \bigg|m_2 \cos^2\uptheta_{13} \sin^2\uptheta_{12} e^{i\alpha_2}
+m_3 \sin^2\uptheta_{13} e^{-2i\delta}\nonumber\\
&&-f_A(M_2) \left[
\sqrt{m_3} \cos\omega \sin\uptheta_{13}  e^{-i\delta}
+\sqrt{m_2} \sin\omega \sin\uptheta_{12} \cos\uptheta_{13}  e^{i\alpha_2/2}
 \right]^2\nonumber\\
&&-f_A(M_1)\left[
-\sqrt{m_3} \sin\omega \sin\uptheta_{13}  e^{-i\delta}
+\sqrt{m_2} \cos\omega \sin\uptheta_{12} \cos\uptheta_{13}  e^{i\alpha_2/2}
\right]^2\bigg|
\end{eqnarray}
for normal ordering and
\begin{eqnarray}
m_{\beta\beta}&=&
\cos^2\uptheta_{13} 
\bigg|
m_1 e^{i\alpha_1}
\cos^2\uptheta_{12}
+ m_2 e^{i\alpha_2}\sin^2\uptheta_{12}\nonumber\\
&&- f_A(M_2)\left[
e^{i\alpha_2/2}\sqrt{m_2} \cos\omega \sin\uptheta_{12} 
+ e^{i\alpha_1/2}\sqrt{m_1} \sin\omega \cos\uptheta_{12}
\right]^2\nonumber\\
&&- f_A(M_1)\left[
-e^{i\alpha_2/2}\sqrt{m_2} \sin\omega \sin\uptheta_{12} 
+ e^{i\alpha_1/2}\sqrt{m_1} \cos\omega \cos\uptheta_{12}
\right]^2
\bigg|
\end{eqnarray}
for inverted ordering.
For $n=2$, it is convenient to choose 
\begin{equation}
\bar{M}=\frac{M_2 + M_1}{2}
\end{equation}
and define
\begin{equation}
\Delta M = \frac{M_2 - M_1}{2}. 
\end{equation}
Since leptogenesis with $n=2$ requires a mass degeneracy, $\bar{M}$ in this case has a physical meaning as the common mass of the heavy neutrinos.
This allows to express Eq.~(\ref{meerewritten}) as 
\begin{eqnarray}
m_{\beta\beta}\simeq \left|
[1-f_A(\bar{M})]m_{\beta\beta}^\nu
+ 2 f_A^2(\bar{M})\frac{\bar{M}^2}{\Lambda^2}\Delta M \left(
\Theta_{e1}^2 - \Theta_{e2}^2
\right)
\right|,
\end{eqnarray}
where we have neglected higher order terms in $\Delta M/\bar{M}$.
In the term that is proportional to $m_{\beta\beta}^\nu$, the contribution from $N_I$ exchange interferes destructively and reduces $m_{\beta\beta}$. The second term can have either sign and can reduce or enhance $m_{\beta\beta}$. The largest effect is expected if the mass splitting $\Delta M$ is relatively large and the mixings $\Theta_{e I}$ of $N_1$ and $N_2$ with the electron flavour are maximally different. Using the fact that the lightest neutrino is massless for $n=2$
($m_{\rm lightest}=0$)
and one of the light neutrino mass splittings is much larger than the other ($\Delta m_{\rm atm}^2\gg \Delta m_{\rm sol}^2$), we can approximate
\begin{align}
{\rm for} \ {\rm NO}: m_{\beta\beta}\simeq &  \bigg|
[1-f_A(\bar{M})]m_{\beta\beta}^\nu
+
2 f_A^2(\bar{M})\frac{\bar{M}^2}{\Lambda^2}
\frac{\Delta M}{\bar{M}}|\Delta m_{\rm atm}|  
e^{-2i\delta} \sin^2\uptheta_{13} \cos(2\omega)
\bigg|, \\
{\rm for} \ {\rm IO}: m_{\beta\beta}\simeq & \bigg|
[1-f_A(\bar{M})] m_{\beta\beta}^\nu
+
2 f_A^2(\bar{M})\frac{\bar{M}^2}{\Lambda^2}
\frac{\Delta M}{\bar{M}}
|\Delta m_{\rm atm}|\cos^2\uptheta_{13}  \\
&\times\Big[\left(e^{i\alpha_2}\sin^2\uptheta_{12}
-e^{i\alpha_1}\cos^2\uptheta_{12}
\right)\cos(2\omega)
+e^{i(\alpha_1 + \alpha_2)/2}\xi\sin(2\uptheta_{12})\sin(2\omega)\nonumber
\Big]
\bigg|.
\end{align}
This shows that, for given $\bar{M}$ and $\Delta M$, one can in principle make the  term proportional to $\Delta M$ arbitrarily large by choosing a sufficiently large $|{\rm Im}\omega|$. In the limit ${\rm Im}\omega\gg 1$ one finds
\begin{align}
{\rm for} \ {\rm NO}: m_{\beta\beta}\simeq &  \bigg|
[1-f_A(\bar{M})]m_{\beta\beta}^\nu\\
&+
 f_A^2(\bar{M})\frac{\bar{M}^2}{\Lambda^2}
\frac{\Delta M}{\bar{M}}|\Delta m_{\rm atm}|  \sin^2\uptheta_{13} 
e^{2{\rm Im}\omega}e^{-2i({\rm Re}\omega + \delta)} 
\bigg|, \nonumber\\
{\rm for} \ {\rm IO}: m_{\beta\beta}\simeq & \bigg|
[1-f_A(\bar{M})] m_{\beta\beta}^\nu\\
&+
 f_A^2(\bar{M})\frac{\bar{M}^2}{\Lambda^2}
\frac{\Delta M}{\bar{M}}
|\Delta m_{\rm atm}|\cos^2\uptheta_{13}  
e^{2{\rm Im}\omega}e^{-2i{\rm Re}\omega} 
\left(\xi e^{i\alpha_2/2}\sin\uptheta_{12}
+ i e^{i\alpha_1/2}\cos\uptheta_{12}
\right)^2
\bigg|.\nonumber
\end{align}
Consistency with neutrino oscillation data at tree level is guaranteed by the use of the Casas Ibarra parameterisation. However, for masses in the GeV range, there exist various constraints on $\Theta_{e I}$ from direct searches for $N_I$ particles, indirect tests involving rare processes and precision observables as well as cosmology that impose upper bounds on $|\Theta_{e I}|^2$. 
These are e.g. summarised in Refs.~\cite{Atre:2009rg,Antusch:2014woa,Drewes:2015iva,deGouvea:2015euy,Fernandez-Martinez:2016lgt} and references therein. In the following we use the analysis in Ref.~\cite{Drewes:2015iva} as a basis.

The comparably strong sensitivity of the term involving $\Delta M$ to the shape of the function $f_A$ implies that the observation of neutrinoless double $\beta$ decay in different nuclei can possibly help to obtain information on the fundamental parameters and $L$ violation even if $\Delta M$ is too small to be resolved experimentally in direct searches for heavy neutrinos.

\section{Baryogenesis}

In leptogenesis, a matter-antimatter asymmetry is  generated in the lepton sector and then partly transferred into a baryon number by weak sphalerons \cite{Kuzmin:1985mm}, which violate $B+L$ and conserve $B-L$. Here $B$ is the total baryon number and $L$ is the total SM lepton number.
In the SM, $B$ is conserved at temperatures $T$ below the temperature $T_{\rm sph}\simeq  130$ GeV \cite{D'Onofrio:2014kta} of sphaleron freezeout. 
Hence, the BAU is determined by the lepton asymmetry $L$ at $T=T_{\rm sph}$.
In the framework of the seesaw mechanism, RH neutrinos with GeV masses must have Yukawa couplings smaller than that of the electron to be consistent with the smallness of the observed neutrino masses and constraints from experimental searches \cite{Drewes:2015iva}. As a result, they may not reach thermal equilibrium in the early universe before $T=T_{\rm sph}$, and the BAU is generated via CP violating flavour oscillations amongst the $N_I$ during their production \cite{Akhmedov:1998qx}.\footnote{An alternative mechanism with $M_I$ in the GeV range has been proposed in Ref.~\cite{Hambye:2016sby}.} 
Since the $N_I$ are highly relativistic at $T>T_{\rm sph}$, the violation of $L$ during this process by the Majorana masses is suppressed as $\sim M_I^2/T^2$. However, sizable asymmetries $L_\alpha$ are generated in the individual flavours $\alpha=e,\mu,\tau$.
These are partly converted into a total $L\neq 0$\footnote{$L$ here refers to the SM lepton number. One can define a generalised lepton number that includes the helicity odd occupation numbers of the heavy neutrino mass eigenstates and remains in good approximation conserved during baryogenesis.}
by a flavour asymmetric washout that hides part of the CP-asymmetry from the sphalerons by storing them in helicity-odd occupation numbers of the $N_I$, which leads to the generation of a $B\neq 0$ by sphalerons. This process crucially relies on the Majorana masses $M_I$ of the heavy neutrinos $N_I$. At the same time, these Majorana masses are responsible for $L$ violation that makes neutrinoless double $\beta$ decay possible in the seesaw model. This immediately raises the question whether the regime in which the $L$ violation due to the masses of heavy neutrinos explain the origin of baryonic matter in the universe may be accessible to neutrinoless double $\beta$ decay searches.
We now study the question whether a value of $m_{\beta\beta}>m_{\beta\beta}^\nu$ can be made consistent with successful leptogenesis via neutrino oscillations in low scale seesaw models. 

\begin{figure}
\begin{center}
\includegraphics[width=0.8\textwidth]{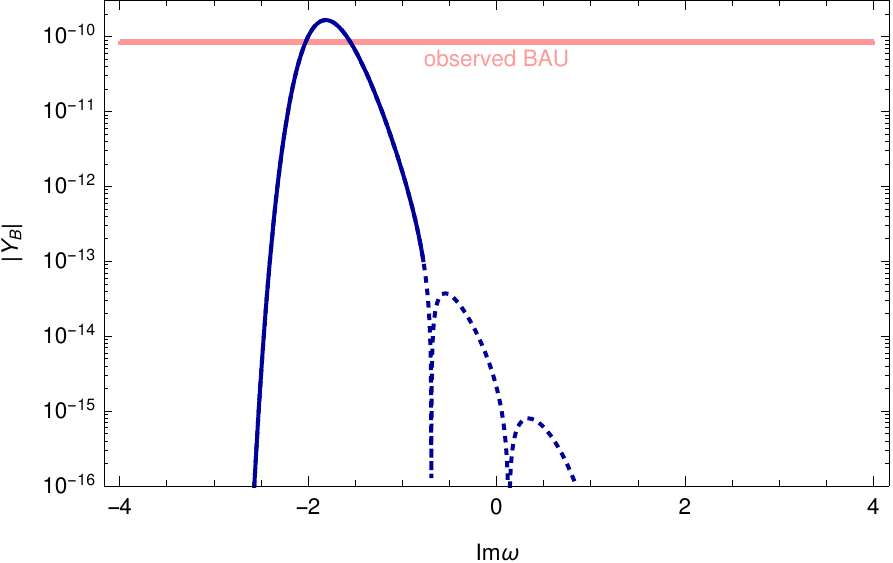}\\
\includegraphics[width=0.8\textwidth]{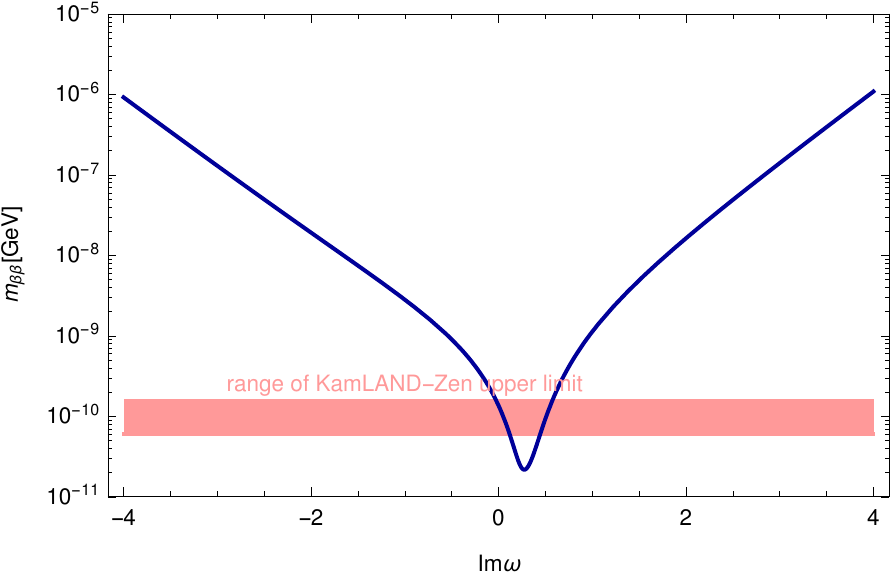}
\end{center}
\caption{
The BAU and $m_{\beta\beta}$ as a function of ${\rm Im}\omega_{13}$.
We fix $M_1=0.22$ GeV, $M_2=0.85$ GeV, $M_3=0.63$ GeV, $m_1=23$ meV, $m_2=24.6$ meV, $m_3=54.6$ meV, $\alpha_1=11.88$, $\alpha_2=11.64$, $\omega_{12}=12.23 + 3.38i$, $\omega_{23}=11.39 - 0.21i$,  $\delta=5.76$ and  
${\rm Re}\omega_{13}=5.18$.
In the dotted region the condition (\ref{conditon2}) is not fulfilled.
Here and in Fig.~\ref{e} we used the radiatively corrected Casas-Ibarra parameterisation introduced in Ref.~\cite{Lopez-Pavon:2015cga} instead of the tree level formula (\ref{CasasIbarraDef}) to ensure consistency with neutrino oscillation data at one loop level.
\label{w}
}
\end{figure}

\begin{figure}
\begin{center}
\includegraphics[width=0.8\textwidth]{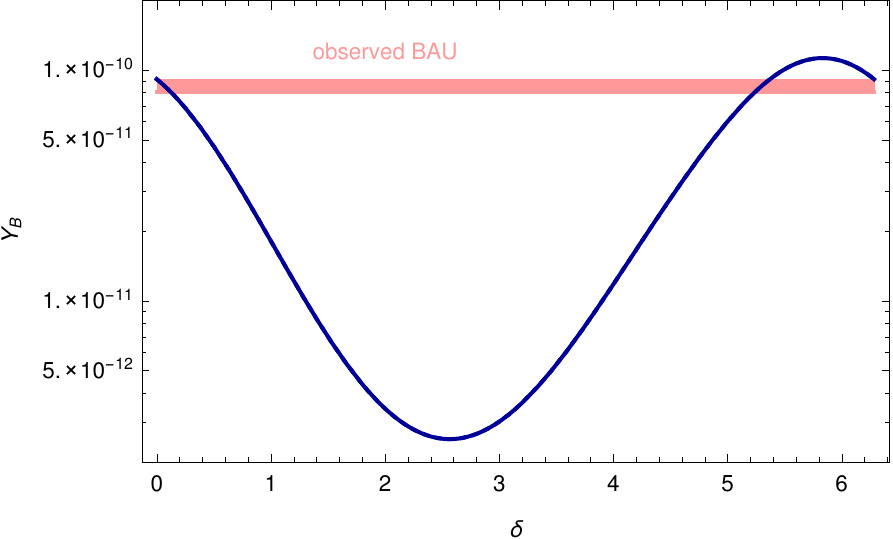}\\
\includegraphics[width=0.8\textwidth]{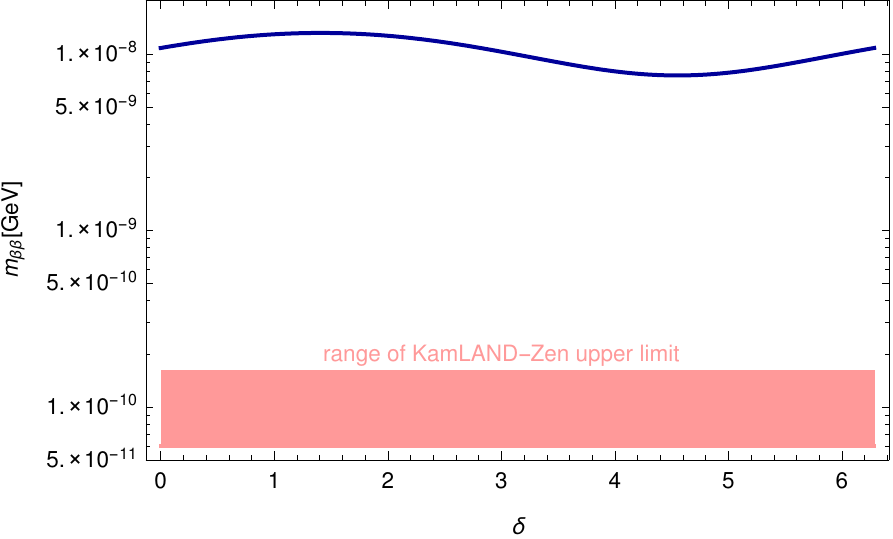}
\end{center}
\caption{
The BAU and $m_{\beta\beta}$ as a function of $\delta$.
We fix $M_1=0.22$ GeV, $M_2=0.85$ GeV, $M_3=0.63$ GeV, $m_1=23$ meV, $m_2=24.6$ meV, $m_3=54.6$ meV, $\alpha_1=11.88$, $\alpha_2=11.64$, $\omega_{12}=12.23 + 3.38i$, $\omega_{23}=11.39 - 0.21i$ and $\omega_{13}=5.18 - 1.62i$.
\label{e}
}
\end{figure}

\paragraph{The case $n=3$} - 
Since a positive contribution to $m_{\beta\beta}$ from $N_I$ exchange can only come from the term $\sum_IM_I\Theta_{e I}^2[f_A(M_I)-f_A(\bar{M})]$ in Eq.~(\ref{meerewritten}), the chances for this are the best in scenarios with $n>2$ that do not require a mass degeneracy. 
However, the parameter space of these scenarios is rather large. Though many authors have studied this process
\cite{Akhmedov:1998qx,Asaka:2005pn,Shaposhnikov:2008pf,Anisimov:2010dk,Anisimov:2010aq,Canetti:2010aw,Garny:2011hg,Garbrecht:2011aw,Canetti:2012vf,Canetti:2012kh,Drewes:2012ma,Canetti:2014dka,Khoze:2013oga,Shuve:2014zua,Garbrecht:2014bfa,Abada:2015rta,Hernandez:2015wna,Kartavtsev:2015vto,Hernandez:2016kel,Drewes:2016gmt}, no complete scan of the parameter space has been performed to date, and such a parameter scan goes beyond the scope of this Letter.
For the sake of a proof of principles, we restrict ourselves to a specific region in the parameter space of the scenario with $n=3$ in which the BAU can be estimated analytically \cite{Canetti:2014dka}.
The rates at which heavy neutrino interaction eigenstates approach thermal equilibrium at temperatures $T\gg M_I$ are governed by the eigenvalues of the matrix $\Gamma_N\simeq F^\dagger F\gamma_{\rm av}T$, c.f. Eq.~(\ref{rates1}), where $\gamma_{\rm av}$ is a numerical coefficient that we set to $\gamma_{\rm av}=0.012$ here, corresponding to the value from Ref.~\cite{Garbrecht:2014bfa} based on Refs.~\cite{Besak:2012qm,Garbrecht:2013urw}. The rate at which they oscillate is determined by the  mass splittings $M_I^2-M_J^2$.
If the CP violating oscillations that generate flavoured asymmetries $Y_\alpha$ occur long before one of the $N_I$ comes into thermal equilibrium, then the generation of the $Y_\alpha$ and the washout (which leads to a $B\neq 0$) can be treated as two separate processes.
The condition for this reads
\begin{eqnarray}\label{conditon1}
\frac{||F^\dagger F||\gamma_{\rm av}a_R^{2/3}}{(M_I^2-M_J^2)^{2/3}}\ll 1,
\end{eqnarray}
where $a_R=m_P(45/(4\pi^3 g_*))^{1/2}=T^2/H$ can be interpreted as the comoving temperature in a radiation dominated universe with Hubble parameter $H$. Here $m_P$ is the Planck mass, $g_*$ the number of degrees of freedom in the primordial plasma and $||F^\dagger F||$ refers to the largest eigenvalue of the matrix. Then the flavoured asymmetries can be estimated as \cite{Drewes:2016gmt}
\begin{eqnarray}
\label{flavoured:asymmetries}
Y_\alpha
&\approx&-\sum\limits_{\overset{I,J,\beta}{I\not=I}}
\frac{{\rm Im}[F_{\alpha I} F_{I\beta}^\dagger F_{\beta J} F_{J\alpha}^\dagger]}{{\rm sign}(M_I^2-M_J^2)}
\left(\frac{m_{\rm Pl}^2}{|M_I^2-M_J^2|}\right)^{\frac 23} 
3.4
\times 10^{-4}
\gamma_{\rm av}^2\
.\end{eqnarray}
Once some heavy neutrino interaction eigenstates approach equilibrium, the washout of the asymmetries $Y_\alpha$ begins. For $T\gg M_I$, the rate for this process is roughly given by $\Gamma^{\alpha}_L\simeq (F F^\dagger)_{\alpha\alpha} \gamma_{\rm av}T/g_w$ with $g_w=2$.\footnote{The factor $g_w$ accounts for the fact that $\gamma_{\rm av}$ has been determined in the context of $\Gamma_N$, which interacts with both components of the SU(2) doublet $\ell_L$, while the $Y_\alpha$ violating interactions of $\ell_L$ only involve the singlet $\nu_R$.}
If two SM flavours come into equilibrium before sphaleron freezeout,\footnote{
If the initial asymmetries $Y_\beta$ in flavours other than $\alpha$ are much larger than $Y_\alpha$, the stronger condition 
$|Y_\alpha e^{-\Gamma^\alpha_L/H}|\gg
|\sum_{\beta\neq\alpha} Y_\beta e^{-\Gamma^\beta_L/H}|$ should be used at $T=T_{EW}$.
}
\begin{equation}\label{conditon2}
\Gamma^{\beta\neq\alpha}_L/H \gg 1  \  {\rm at} \ T=T_{\rm sph},
\end{equation}
then the BAU can be estimated as 
\begin{equation}\label{BAU}
Y_B\simeq -\frac{28}{79}Y_\alpha\frac{3}{7}e^{-\Gamma^\alpha_L/H},
\end{equation}
where $28/79$ is the sphaleron conversion factor, the factor $3/7$ comes from the equilibration of all charges except $Y_\alpha$ during their washout and the exponential describes the washout of $Y_\alpha$ itself.
By plugging numbers into the parametrisation (\ref{CasasIbarraDef}), it is straightforward to see that $m_{\beta\beta}>m_{\beta\beta}^\nu$ can be realised while producing a BAU that exceeds the observed value and respecting the conditions (\ref{conditon1}) and (\ref{conditon2}).
We illustrate the parameter dependence of $m_{\beta\beta}$ and $Y_B$ on the observable Dirac phase $\delta$ and ${\rm Im}\omega_{23}$ in figures \ref{w} and \ref{e} to show that a large $m_{\beta\beta}$ can indeed be realised while explaining the observed BAU. 
The quantities ${\rm Im}\omega_{ij}$ determine the magnitude of the active-sterile mixing $U_{\alpha I}^2$ and can thereby be constrained experimentally if heavy neutrinos are found in the laboratory.
This treatment is of course very simplified and should be understood as a proof of principle. A detailed study of the parameter space in the region where the conditions do not apply requires a numerical solution of the quantum kinetic equations for each point in parameter space. 

\begin{figure}
\begin{center}
\includegraphics[width=0.8\textwidth]{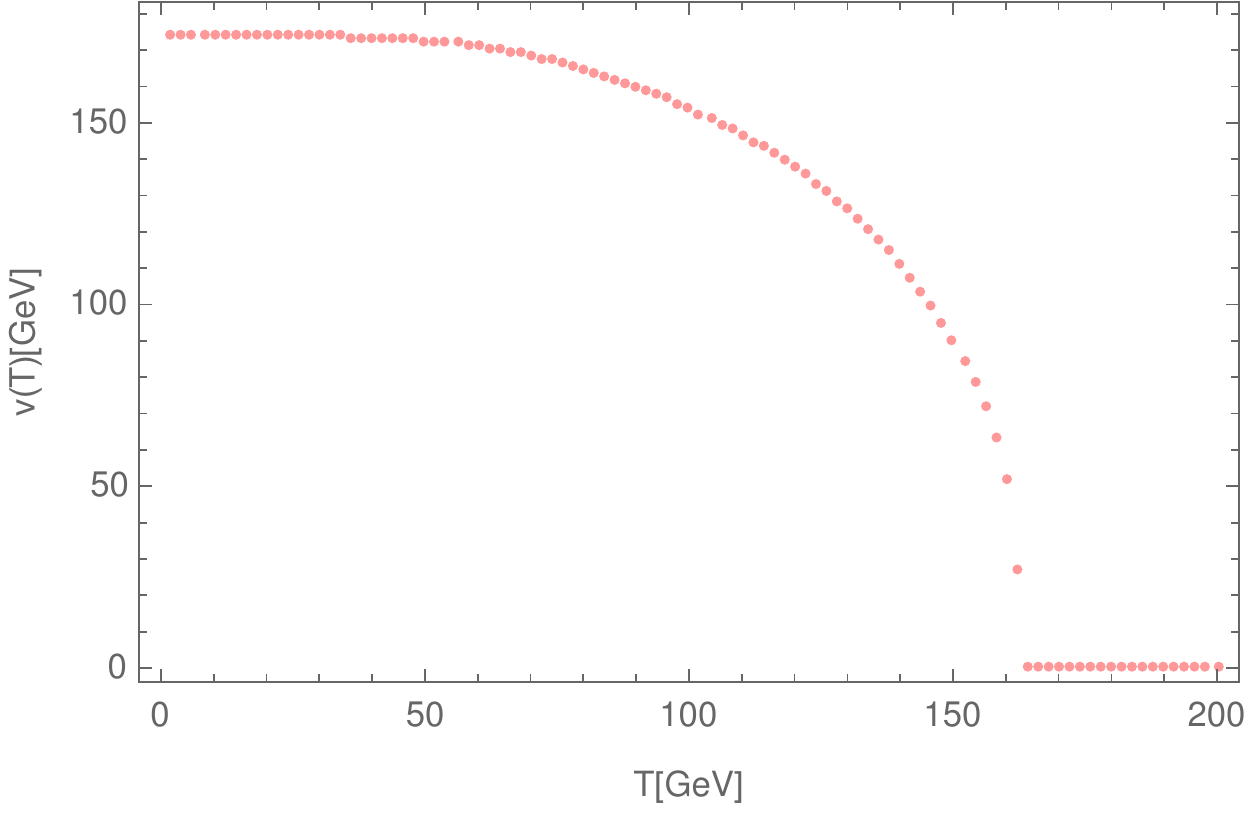}
\end{center}
\caption{
The function $v(T)$ used in our calculation.\label{higgsvev}
}
\end{figure}

\paragraph{The case $n=2$} -
For a more quantitative treatment we return to the scenario with $n=2$, where the lower dimensionality of the parameter space makes a numerical scan less expensive. 
It is well-known that leptogenesis in this scenario requires a mass degeneracy of order $|\Delta M|/\bar{M}\ll 1$ 
\cite{Canetti:2012vf,Canetti:2012kh,Shuve:2014zua,Hernandez:2015wna}. We perform a numerical scan in order to address the question whether successful baryogenesis and $m_{\beta\beta}>m_{\beta\beta}^\nu$ can be realised simultaneously for $n=2$. Phenomenologically this is interesting because this scenario effectively describes baryogenesis in the $\nu$MSM.
In order to identify the parameter region where baryogenesis is possible, we solve momentum integrated kinetic equations for the two helicity components $\rho_{N}$  ans $\rho_{\bar{N}}$ of the heavy neutrino density matrix and $Y_\alpha$ \cite{Asaka:2005pn,Shaposhnikov:2008pf},
\begin{eqnarray}
\label{kinequ1}
i \frac{1}{\mathcal{H}X}\frac{d\rho_{N}}{d X}&=&[H_N, \rho_{N}]-\frac{i}{2}\{\Gamma_N, \rho_{N} - \rho^{eq}\} +\frac{i}{2} Y_\alpha{\tilde\Gamma^\alpha}_N~,\\
i \frac{1}{\mathcal{H}X}\frac{d\rho_{\bar{N}}}{d X}&=& [H_N^*, \rho_{\bar{N}}]-\frac{i}{2}\{\Gamma^*_N, \rho_{\bar{N}} - \rho^{eq}\} -\frac{i}{2} Y_\alpha{\tilde\Gamma^{\alpha *}}_N~,\label{kinequ2}\\
i \frac{1}{\mathcal{H}X}\frac{dY_\alpha}{d X}&=&-i\Gamma^\alpha_L Y_\alpha +
i {\rm tr}\left[{\tilde \Gamma^\alpha}_L(\rho_{N} -\rho^{eq})\right] -
i {\rm tr}\left[{\tilde \Gamma^{\alpha*}}_L(\rho_{\bar{N}}  -\rho^{eq})\right]
~.
\label{kinequ3}
\end{eqnarray}
Here $\rho^{eq}$ is the equilibrium density matrix and $X=\bar{M}/T$ is a dimensionless time variable.  The function
\begin{equation}
\mathcal{H}\equiv -\frac{\partial}{\partial X}\sqrt{\frac{45}{4\pi^3 g_*}}\frac{m_P}{2M^2} X 
\end{equation}
can be identified with the Hubble parameter if the number of degrees of freedom $g_*$ is constant during the evolution, which is justified in the present context.
The coefficients appearing in Eqns.~(\ref{kinequ1})-(\ref{kinequ2}) can be
expressed as
\begin{eqnarray}
H_N&=&\frac{1}{4T}\left[
-2\bar{M}\Delta M \sigma_3 + \tilde{F}^\dagger \tilde{F}\frac{T^2}{4} + \tilde{F}^\dagger \tilde{F} v^2(T)
\right]\\
\Gamma_N&=&\phantom{i} \sum_{\alpha}\big(\tilde{F}^*_{\alpha I}\tilde{F}_{\alpha J}R(T,M)_{\alpha\alpha}+\tilde{F}_{\alpha I}\tilde{F}^*_{\alpha J}R_M(T,M)_{\alpha\alpha}\big),\label{rates1}\\
(\tilde{\Gamma}_L^\alpha)_{IJ} \simeq (\tilde{\Gamma}_N^\alpha)_{IJ}&=& \phantom{i} \big( \tilde{F}^*_{\alpha I}\tilde{F}_{\alpha J}R(T,M)_{\alpha\alpha}-\tilde{F}_{\alpha I}\tilde{F}^*_{\alpha J}R_M(T,M)_{\alpha\alpha}\big),\label{rates2} \\
\Gamma_L^\alpha
&=&\phantom{i} \frac{1}{g_w}\big((FF^\dagger)_{\alpha\alpha}\left(R(T,M)_{\alpha\alpha}+R_M(T,M)_{\alpha\alpha}\right)\big)\label{rates3}, 
\end{eqnarray}
with $\tilde{F}=FU_N\simeq F$.
The function $v(T)$ is shown in figure \ref{higgsvev}.
We have assumed that the average momentum of heavy neutrinos is $|\p|\simeq 2T$.
In the limit $T\gg M_I$ one can approximate $R_M\simeq0$, $R\simeq \gamma_{\rm av} T$.
The equations (\ref{kinequ1})-(\ref{kinequ3}) are the heavy neutrino equivalent of the density matrix equations commonly used in neutrino physics \cite{Sigl:1992fn} and are derived in the appendix of Ref.~\cite{Canetti:2012kh}.
Our scan comprises $5\times 10^7$ parameter choices for each neutrino mass ordering.
We use a logarithmic prior for the mass splitting in the interval $-16 \leq \log(\Delta M/{\rm GeV})/\log10 \leq 0$  and flat priors in all other quantities in the the parametrisation (\ref{CasasIbarraDef}). We considere the mass range $0.1 {\rm GeV} < \bar{M} < 5$ GeV. 
We accept a point when the generated BAU lies within a $5\sigma$ range of the observed value $\eta_B=(8.06 - 9.11)\times 10^{-11}$ \cite{Ade:2015xua}.
At the same time, we require consistency with all direct and indirect constraints on the low scale seesaw that are summarised in Ref.~\cite{Drewes:2015iva} (except the constraint on $m_{\beta\beta}$ of course).
These include indirect experimental constraints from neutrino oscillation data, electroweak precision data, lepton universality, searches for rare lepton decays and tests of CKM unitarity with bounds from big bang nucleosynthesis and past direct searches at colliders and fixed target experiments. 

\begin{figure}
\begin{center}
\includegraphics[width=0.8\textwidth]{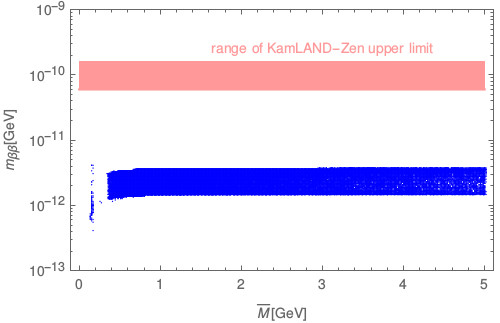}\\
\includegraphics[width=0.8\textwidth]{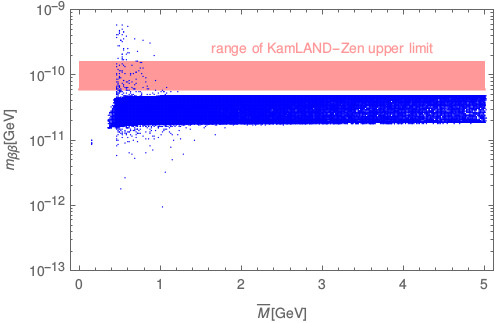}
\end{center}
\caption{\label{scanfig}
The blue points correspond to 
 values of $\bar{M}$ and $m_{\beta\beta}$ 
that are consistent with successful leptogenesis and the constraints on the low scale seesaw summarised in Ref.~\cite{Drewes:2015iva}. The red band shows the upper limit on $m_{\beta\beta}$ from the KamLAND-Zen experiment \cite{KamLAND-Zen:2016pfg}, where  the width of the band comes from the theoretical uncertainty in the nuclear matrix elements that affects the translation from a bound on the lifetime into a bound on $m_{\beta\beta}$. 
The upper plot is for normal mass ordering, the lower for inverted mass ordering.
}
\end{figure}

\begin{figure}
\begin{center}
\includegraphics[width=0.8\textwidth]{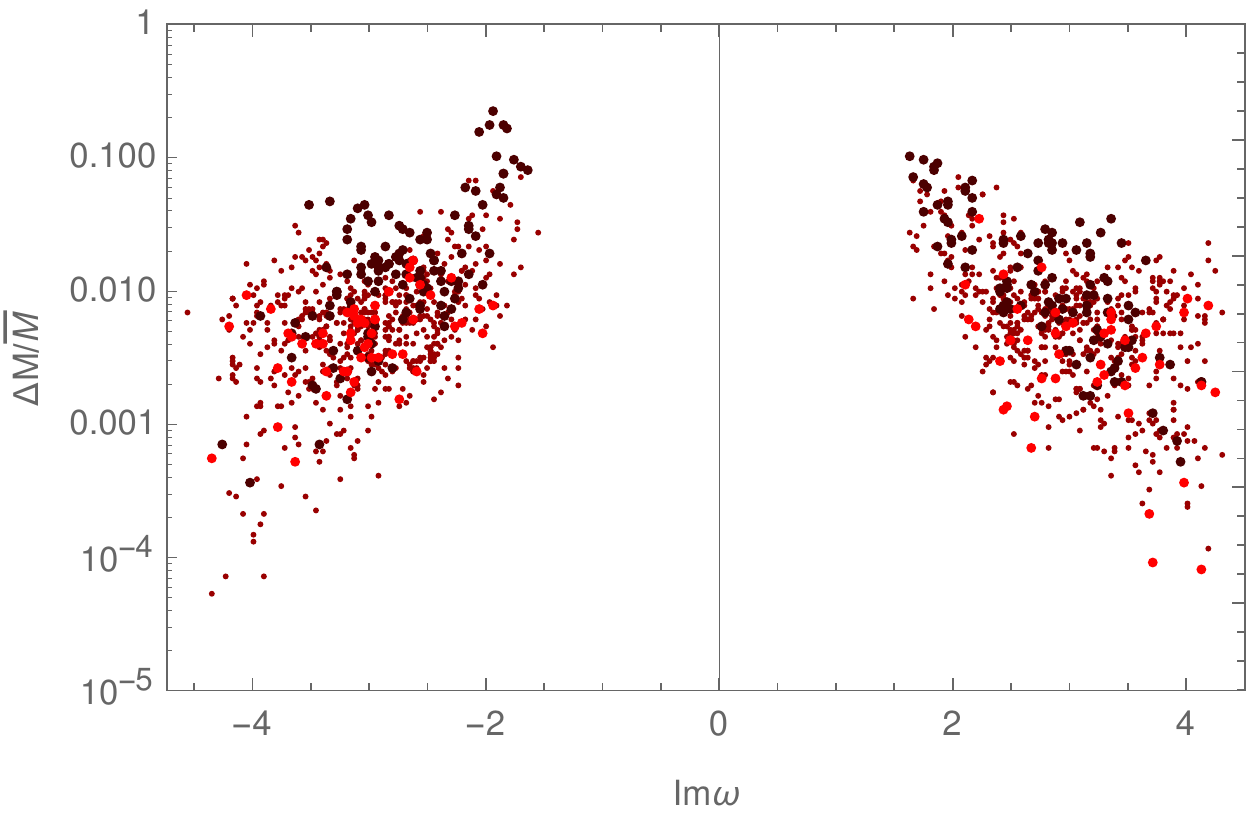}
\end{center}
\caption{\label{ImOmegaDeltaMplot}
A representative distribution of parameter values that lead to successful baryogenesis and $m_{\beta\beta}>m_{\beta\beta}^\nu$ while being in agreement with all other direct and indirect constraints discussed in Ref.~\cite{Drewes:2015iva}.
The colour indicates the magnitude of $\bar{M}$, which ranges from values below the kaon mass (lightest) to values above the D-meson mass (darkest). 
}
\end{figure}

The result of this scan is shown in figure \ref{scanfig}.
The densely populated area corresponds to the standard prediction $m_{\beta\beta}^\nu$. For $\bar{M}>2$ GeV we find almost no points outside this region because the suppression of the heavy neutrino contribution due to $f_A$ is efficient. For lower masses, we find deviations from the standard prediction in both directions. For inverted ordering the value of $m_{\beta\beta}$ can exceed the present day experimental limit from the KamLAND-Zen \cite{KamLAND-Zen:2016pfg} and GERDA \cite{Agostini:2013mzu} experiments.
This implies that neutrinoless double $\beta$ decay experiments have already started to rule out part of the leptogenesis parameter space that is not constrained by any other experiment.
The allowed parameter region with $m_{\beta\beta}>m_{\beta\beta}^\nu$ is characterised by relatively large mass splitting and large $|{\rm Im}\omega|$, e.g. $\Delta M/\bar{M}\sim 10^{-3}$ and $|{\rm Im}\omega|>2$, see Fig.~\ref{ImOmegaDeltaMplot}. \footnote{These results agree with what was found in the analyses in Refs.~\cite{Asaka:2016zib,Hernandez:2016kel}, which were performed in parallel to our analysis and appeared on arxiv.org in the same week.
The main results of Ref.~\cite{Hernandez:2016kel} had been presented by Pilar Hernandez at the MIAPP workshop \emph{Why is there more Matter than Antimatter in the Universe?} the week before.}
To the best of our knowledge, this parameter region is not singled out by any known symmetry, which seems to imply that a large value of $m_{\beta\beta}$ for $n=2$ requires considerable tuning.
For $\bar{M}$  below the kaon mass the viable parameter space rapidly shrinks because $|\Theta_{\alpha I}|^2$ is constrained from below by the requirement that the $N_I$ decay before BBN and constrained from above by direct searches in fixed target experiments.

\section{Conclusions}
We conclude that the rate of neutrinoless double $\beta$ decay in low scale leptogenesis scenarios within the minimal seesaw model with Majorana masses in the GeV range  
can be both, smaller and larger than the expectation from light neutrino exchange alone, while respecting all known constraints on the properties of heavy neutrinos from experiments and cosmology. 
For inverted ordering the value of $m_{\beta\beta}$ can exceed the present day experimental limit, which implies that neutrinoless double $\beta$ decay experiments have already started to rule out part of the leptogenesis parameter space that is not constrained by any other experiment.
The observation of a value of $m_{\beta\beta}$ that deviates from the standard prediction would contain valuable information about the heavy neutrino mass splitting and the CP-violating phases in their couplings. Together with a measurement of the Dirac phase $\delta$ in neutrino oscillation experiments, this would allow to impose strong constraints on the violation of lepton number and CP in the low scale seesaw model.
If any heavy neutral leptons are discovered in future experiments and their mixings $|\Theta_{\alpha I}|^2$ with the SM neutrinos have been measured, this information will be crucial to decide whether theses particles are indeed responsible for the generation of baryonic matter in the universe.\\ \\ 


\section*{Acknowledgements}
We would like to thank Mikhail Shaposhnikov for helpful discussions in the initial phase of this project and for sponsoring MaD's visit to Lausanne that made this project possible.
We would also like to than Fedor Bezrukov for his comments on the final version of this manuscript.
This work was supported by the Deutsche Forschungsgemeinschaft (DFG) and the Swiss National Science Foundation (SNF).


\bibliography{all}

\end{document}